\documentclass[pre,twocolumn,aps,showpacs]{revtex4}
\usepackage{graphicx}

\begin{document}

\author{I.M. Sokolov}
\affiliation{%
Institut f\"{u}r Physik, Humboldt-Universit\"{a}t zu Berlin,
Invalidenstr. 110, D-10115 Berlin, Germany}
\title{Optimizing a Ratchet Gear}
\date{\today}

\begin{abstract}
The energetic efficiencies of rocked ratchets reported in the literature
typically lie in the sub-percent range. We discuss the problem of
optimization of the energetic efficiency of a ratchet, and show that
considerably higher efficiencies can be achieved; however this assumes a
fine-tuning of the parameters of the system. The domain of parameters
corresponding to high efficiencies is typically narrow.
\end{abstract}


\maketitle

The recent interest to thermodynamics of ratchet devices, inspired by
biological applications (where ratchets serve as gears of molecular motors
powering cells and subcell units, see Ref. \cite{Frey} for an introductory
review), stays strong for already a decade. The simplest model
(oversimplified compared to any biological system, but physically still not
elementary) is a so-called rocked ratchet. The model corresponds to a
particle moving in a spatially asymmetric potential under the influence of
an external field, either periodic or stochastic, of strong friction and of
thermal noise. This model was one of the first ones discussed by physicists 
\cite{Magnasco} and is investigated to a great detail, see Refs.\cite
{J+A+P,Reimann,R+H} for comprehensive reviews. The energetic efficiency is
one of the simplest (and deepest) thermodynamic characteristics of such
systems and is now under extensive investigation, see Ref.\cite{P+C} for a
review. Even if this efficiency is not a crucial parameter in biological
systems or in nanomechanical appliances, it is still of primary importance
since it determines the heat production under operation and thus the overall
heat regime. The energetic efficiency of a rocked ratchet is notoriously
low: The Refs. \cite{Takagi1,Sumithra} discussing this issue give
numerical values of the efficiencies in a sub-percent domain; Ref. \cite{SB1}
discussing a related discrete model presents similarly low values. On the
other hand, the values of parameters of the ratchets discussed in these
works are arbitrary, so that these efficiencies may be low just by chance.
In what follows we discuss this issue in detail and show that rather high
efficiencies can be attained; however the domain of parameters, where the
efficiencies are high, is narrow. Thus, a ratchet, as a technical device,
needs scrupulous optimization if the efficient performance is aimed.

Let us first discuss the parameters of the ratchet gear. A rocked ratchet is
defined by a Langevin equation,

\begin{equation}
\dot{x}=\mu \left[ F(x)+f(x,t)\right] +\xi (t),  \label{Lang}
\end{equation}
where $\mu $ is the mobility of the particles, and $\xi (t)$ is a $\delta $%
-correlated Gaussian Langevin force with zero mean and with $\overline{\xi
^{2}(t)}=2\theta \mu $, where $\theta $ is the energetic temperature. Here $%
F(x)$ is a force corresponding to the ratchet potential and $%
f(x,t)=f_{0}+f_{1}(t)$ is a sum of the external see-saw force $f_{1}(t)$
with zero mean and of the constant force $f_{0}$ against which the useful
work is done by pumping particles uphill. The temporal evolution of the
particles' distribution is given by a Fokker-Planck equation, 
\begin{equation}
\frac{\partial p(x,t)}{\partial t}=\frac{\partial }{\partial x}\left( D\frac{%
\partial p(x,t)}{\partial x}+\mu p(x,t)\frac{\partial }{\partial x}%
U(x,t)\right) ,  \label{FoPla}
\end{equation}
where $U(x)=V(x)+f(t)x$ is the overall potential, and $D=\theta \mu $ is the
diffusion coefficient. In our work, we take $F(x)$ to be piecewise-constant,
so that the potential $V(x)$ is a saw-tooth function: 
\begin{equation}
V(x)=\left\{ 
\begin{array}{ll}
Vx/a & \text{for }0<x\leq a \\ 
V(L-x)/(L-a) & \text{for }a<x\leq L
\end{array}
.\right.  \label{rat}
\end{equation}
The mobility $\mu $ is set to unity. Apart from the time-dependence
(protocol) of the external force, which is a \textit{function}, the
parameters characterizing the situation are: the geometrical parameters $L$
and $a$, characterizing the ratchet, the amplitude of the saw-tooth
potential $V$, the energetic temperature $\theta $, and value $f_{0}$ of a
constant force against which the work is done. The geometrical parameters
and $V$ characterize the appliance itself, $\theta $ characterizes the
external conditions, and $f_{0}$ will be tuned in order to achieve the
maximal efficiency under other conditions fixed. Although some parameters
can be absorbed into dimensionless combinations, their overall number is
still too large to allow for simple optimization. In what follows we will
use the same ratchet as in Refs. \cite{Takagi1} and \cite{Sumithra}, which
in our notation corresponds to $L=1,$ $a=0.8$ and $V=1$. Moreover, in order
to be able to compare the results, we use the same protocol of the external
force, switching between the values $+f_{1}$ and $-f_{1}$, and having a
period $T$. \ The overall force $f(t)$ meanders in this case between the
values $f_{-}=f_{0}-f_{1}$ and $f_{+}=f_{0}+f_{1}$. Under homogeneous in
space forcing field and load force, the energetic efficiency $\eta =W/A$
(where $W$ is the useful work and $A$ is the input work) is given by \cite
{Takagi1,Sokolov3}

\begin{equation}
\eta =-\frac{\overline{I(f_{0}+f_{1})f_{0}}}{\overline{I(f_{0}+f_{1})f_{1}}},
\label{Effi}
\end{equation}
where $I$ is the overall current through the system, and the mean values are
taken over the period of forcing.

Let us first discuss the efficiency under the adiabatic mode of the
operation, just as it was done in Ref. \cite{Takagi1}, so that $T\rightarrow
\infty $. Now, only 3 parameters are left: the amplitude $f$ of the external
forcing, the force $f_{0}$ (external load), and the temperature $\theta $.
Under a very slowly changing force, the system can simply be described as a
rectifier (nonlinear element) whose ''Volt-Ampere'' characteristics can
easily be calculated in the adiabatic approximation. The current $I(f,\theta )
$ is the given by \cite{Sokolov3} 
\begin{eqnarray}
&&\mu I^{-1}=\frac{a}{f-V/a}+\frac{L-a}{f+V/(L-a)}+ \label{Curr} \\
&&\theta \left\{ \frac{\exp \left( -\frac{V-fa}{\theta }\right) -1}{%
(f-V/a)^{2}}-\frac{\exp \left( \frac{f(L-a)+V}{\theta }\right) -1}{\left[
f+V/(L-a)\right] ^{2}}\right\} + \nonumber \\
&&\theta \left[ \frac{\exp \left( -\frac{V-fa}{\theta }\right) -1}{f-V/a}-%
\frac{\exp \left( \frac{f(L-a)+V}{\theta }\right) -1}{f+V/(L-a)}\right]
^{2}\times  \nonumber \\
&&\left[ \exp \left( \frac{f(L-a)+V}{\theta }\right) -\exp \left( -\frac{V-fa}{%
\theta }\right) \right] ^{-1} .  \nonumber
\end{eqnarray}
From this expression the limiting forms for $\theta \rightarrow \infty $ and
for $\theta \rightarrow 0$ readily follow. Thus for $\theta \rightarrow
\infty $ one has $I=\mu f+O(\theta ^{-2})$: the nonlinearities vanish, and
the rectifier doesn't work, on the other hand, for $\theta \rightarrow 0$
the current is given by 
\begin{equation}
I(f)=\left\{ 
\begin{array}{l}
0\qquad \text{for }-V/(L-a)<f<V/a \\ 
\mu \left( \frac{a}{f-V/a}+\frac{L-a}{f+V/(L-a)}\right) ^{-1} \text{otherwise}
\end{array}
.\right.   \label{Deter}
\end{equation}
The interval $-V/(L-a)<f<V/a$ where the current is zero is termed the
mobility gap. This expression corresponds to the current in the
deterministic mode of operation, which was discussed in detail in Ref.\cite
{Sokolov3}. We use this expression to plot the efficiency as a function of $%
f_{-}$ and $f_{+}$ in upper panel of Fig.1. Here only positive
values of efficiencies are plotted; the efficiency in the regimes where the
useful work is negative, is set to zero. According to Ref.\cite{Sokolov3}
where the optimization of a deterministic ratchet was discussed, the maximal
efficiency in the deterministic regime under symmetric forcing depends only
on the ratchet's geometry and is given by 
\begin{equation}
\eta _{\mathrm{sym}}=\left| 1-2a/L\right| .  \label{max}
\end{equation}
The maximum is achieved under 
\begin{equation}
f_{0,\max }=-\frac{V(L-2a)}{2a(L-a)},\text{ }\left| f_{1,\max }\right| =%
\frac{VL}{2a(L-a)},  \label{Opt}
\end{equation}
i.e. under $f_{-}=-V/(L-a)$ and $f_{+}=V/a$, corresponding to the mobility
thresholds.

\begin{figure}
\scalebox{0.77}{\includegraphics{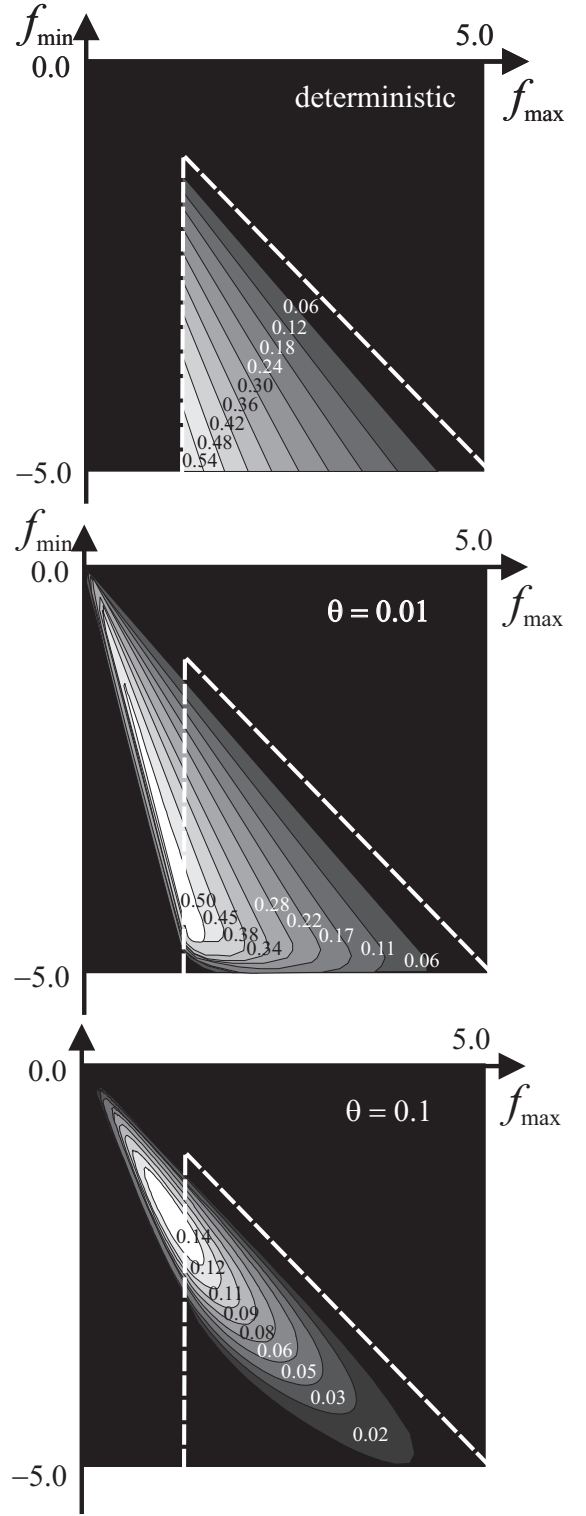}}
\caption{Contour plots of efficiencies of a rocked ratchet
as functions of $f_{-}$ and $f_{+}$. The upper panel 
corresponds to the deterministic mode of operation, the middle panel
presents the results for a thermal ratchet at $\theta=0.01$, and the lower
panel corresponds to  $\theta=0.1$. The dashed lines denote the boundary 
of the region in which the deterministic ratchet produces positive work.}
\end{figure}

Note that the maximal efficiency under adiabatic forcing in the
deterministic regime for the ratchet considered in Ref. \cite{Takagi1} would
be $\eta _{\max }=0.6$, three orders of magnitude higher than the efficiency
reported in this work even for very low temperatures.

\begin{figure}
\scalebox{0.45}{\includegraphics{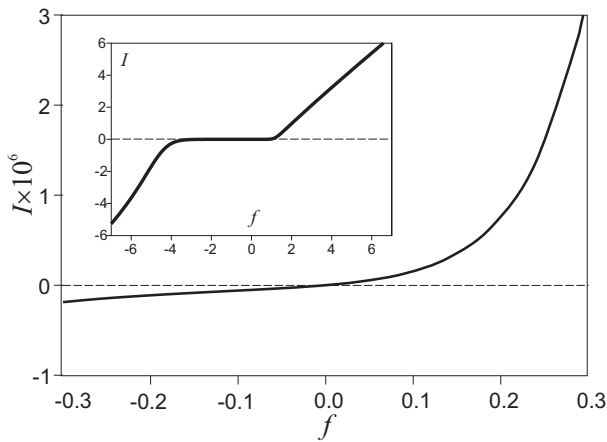}}
\caption{This figure illustrates the strong nonlinearity of a thermal
ratchet in the domain corresponding to the mobility gap of
its deterministic counterpart. The inset shows the overall load-current
characteristic of a system. Here $\theta=0.01$, see text for details.}
\end{figure}

In the middle panel of Fig.1 we plot the efficiency $\eta $ at very low
energetic temperature $\theta =0.01$, where now the expression, Eq.(\ref
{Curr}) for the current is used. One readily infers, that the maximal
efficiency didn't decrease strongly (now one has $\eta _{\max }=0.56$), and
moreover that the overall region of parameters, in which positive
efficiencies are achieved is considerably broadened! A white triangle in the
left panel just denotes the domain of parameters in which the efficiency
under deterministic operation is positive. The part of lying outside of
this triangle where the positive efficiencies are still achieved,
corresponds to $f_{-}$ and $f_{+}$ within the mobility gap. The
explanation of the effect is very simple: in the gap, the current (as a
function of the force) vanishes as $\theta \rightarrow 0$ is considered, thus,
for low temperatures, this current is very small. This is the domain where
the deterministic ratchet doesn't work at all. On the other hand, since one
finds this current in both averages in the enumerator and in the denominator
of Eq.(\ref{Effi}), the quotient does not have to be small or even to
vanish. Indeed, it tends to a constant value, due to the fact that at low
but finite temperatures the dependence of $I(f)$ shows an extremely strong
nonlinearity, see Fig.2. Let us stress this interesting finding: The
deterministic mode of the operation can be considered as a limit of zero
temperature when the currents through the system are considered. However,
the efficiency of a thermal ratchet does not vanish in the domain of
parameters where the current does; so that the thermal ratchet has a larger
domain of high efficiencies than its deterministic counterpart.

The overall dependence of $\eta _{\max }$ on $\theta $ is shown in Fig. 3
for $0\leq \theta \leq 0.2$. We see that in the adiabatic regime it is a
monotonous function of temperature, as discussed in Ref.\cite{Takagi1}. The
numerical results for higher temperatures are less reliable, since the
maxima get to be very sharp. Note that for $\theta =0.1$ the maximal
efficiency under adiabatic operation is still around $\eta _{\max }=0.154$
and that, according to Fig.2 maximal efficiency is achieved for the values
of $f_{-}$ and $f_{+}$ lying near the boundary of the domain where the
deterministic ratchet produces positive work. 

\begin{figure}
\scalebox{0.5}{\includegraphics{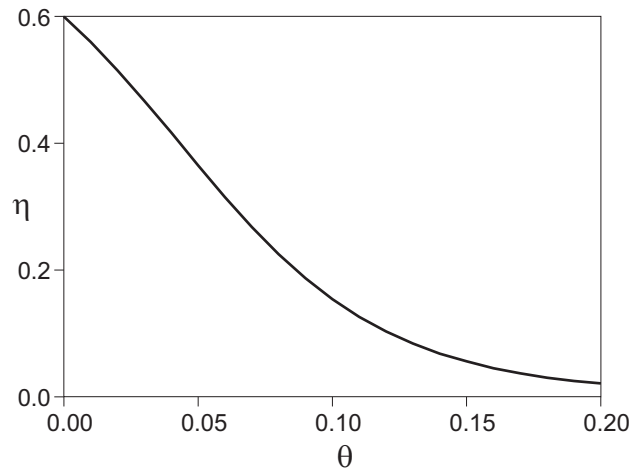}}
\caption{The maximal energetic efficiency of the ratchet as a function of
the energetic temperature $\theta$, see text for details.}
\end{figure}

We also note that the adiabatic ratchet works irreversibly, even under the
limiting transition $a\rightarrow L$, when $\eta \rightarrow 1$ \cite
{Sokolov3}. Such a ratchet, in which one side of the saw-tooth is very
steep, is essentially an ideal rectifier. It performs extremely 
good in a broad temperature range, due to
the possibility of using strongly negative $f_{-}$, but still loses
efficiency for $\theta \geq V$. The reason of the irreversibility is easy to
understand when following the particle's path along the ratchet: for the
positive overall force the particle moves infinitesimally slow when sliding
along the flatter side of the saw-tooth and falls with a constant velocity
along the steeper side, thus producing heat and losses.

So far we have shown that adiabatic ratchets may perform much better than
supposed, provided the temperatures are low enough. In Ref.\cite{Sokolov3}
it is shown, that each deterministic ratchet can also work reversibly, thus
achieving the efficiency $\eta =1$ under the mode of operation which is
quasistatic but not adiabatic. Thus, the reversible regime under the
deterministic mode of operation implies the synchronization of the external
force and particle's position: the external force must change its sign when
the particle passes the apex of the potential, so that the particle's
velocity stays infinitesimally small. In a piecewise-linear potential, Eq.(%
\ref{rat}), and for piecewise-constant, symmetric force $f_{1}(t)$ of period 
$T$ this corresponds to such a choice of the forces that particle passes the
distances $a$ and $L-a$, during the respective half-periods. The velocities
thus are: $v_{1}=2a/T$ and $v_{2}=2(L-a)/T$. This gives us the optimal
values of the forces 
\begin{eqnarray}
&&f_{0,\max }=-\frac{V(2a-L)}{2a(L-a)}+\frac{L}{\mu T},\\ \nonumber
&&\left| f_{1,\max
}\right| =\frac{VL}{2a(L-a)}+\frac{2a-L}{\mu T},
\end{eqnarray}
and the maximal efficiency corresponding to 
\begin{equation}
\eta _{\max }=\frac{-f_{0,\max }L}{-f_{0,\max }L+2\left[
a^{2}+(L-a)^{2}\right] /\mu T}.
\end{equation}

Note that the possibility of the reversible mode of operation is closely
connected to the fine time-tuning of the external force, and that these high
efficiencies follow as a kind of a non-linear resonance: even slight changes
of temporal properties of the force (keeping the amplitudes constant) lead
to a dramatic drop in efficiency. As an example let us consider our system
with the values of forces which optimize the efficiency for $T=40$, which
are $f_{0}=-1.85$, $f_{1}=3.14$ and correspond to $\eta =0.982$. Taking now
a period $T=40.04$ (relative detuning from the ''resonance'' of the order of
10$^{-3}$) we get (through the numerical solution of Eq.(\ref{Lang}) with $%
\xi =0$ and numerical evaluation of $\eta $ using Eq.(\ref{Effi}) and
time-averaging over 10$^{4}$ periods of the field) that the value of $\eta $
drops to $0.60$, only slightly higher that the value corresponding to the
adiabatic operation, which for this values of the forces is $\eta =0.578$.
To understand the situation it is enough to examine the trajectories of the
particle's motion: even at tiny detuning trajectories develop parts
corresponding to the particle's falling down the steep potentials, which
correspond to high losses. The same dramatic drops to the values which are
only slightly higher than the adiabatic ones are seen when increasing the
temperature (switching of $\xi =\sqrt{2\theta }$). 

The active
synchronization (''the Humphrey Potter's solution''), where the see-saw
force is triggered when the particle actually passes the apex \cite{Sokolov3}%
, helps to save the situation and stabilize high efficiencies: posing the
''switches'', say, at distance $\delta $ to the left from the bottom of the
potential and to the right of its maximum, one can achieve the efficiency of 
$\eta =0.961$ for $\delta =10^{-4}$ and $\eta =0.790$ for $\delta =10^{-2}$
under deterministic regime, but the drops in efficiency when increasing
temperature are still dramatic. The effective regimes close to reversibility
seem not to survive under elevated temperatures.

We note that ratchets in a finite-time mode of operation can show the
increase of efficiency with increasing temperature, Ref.\cite{Sumithra,SB1}.
No such regimes were found in the vicinity of the modes of operation,
corresponding to high efficiencies (either reversible or irreversible).
Probably, such situations are pertinent to the outer regions of the domain
where the work is positive, where the overall efficiency is low.

Let us summarize our findings. The efficiencies of the rocked ratchets
reported in the literature are notoriously low; however, since the
parameters of the systems discussed are chosen at random, this could only
mean that the domain of parameters corresponding to high efficiencies is
rather narrow. We show that it is indeed the case, and that the efficiencies
of the rocked ratchets at moderate temperatures can be reasonably high.

The author is indebted to Prof. P. H\"{a}nggi for useful discussions, and to
the Fonds der Chemischen Industrie for partial financial support.

\end{document}